\documentclass[a4paper,11pt]{article}
\usepackage{pos}

\newcommand{\apj}{Astrophysical Journal}
\newcommand{\apjl}{Astrophysical Journal Letters}

\newcommand{\apjs}{Astrophysical Journal Supplement}
\newcommand{\aap}{Astronomy and Astrophysics}

\newcommand{\aapr}{Astronomy and Astrophysics Reviews}
\newcommand{\mnras}{Monthly Notices of the RAS}
\newcommand{\pasa}{Publications of the Astron. Soc. of Australia}
\newcommand{\araa}{Annual Review of Astron and Astrophys}

\newcommand{\nat}{Nature}

\manuallySeparateAuthors

\title{Hunting the gamma-ray emission from Fast Radio Burst with Fermi-LAT}
 \ShortTitle{Gamma-ray emission from FRBs with Fermi-LAT}

\author*[a,b,c]{Giacomo Principe}
\author[d]{, Nicola Omodei}
\author[d]{, Niccolò Di Lalla}
\author[e]{, Leonardo Di Venere}
\author[a,b]{, Francesco Longo}
\author{, on behalf of the Fermi Large Area Telescope Collaboration.}

\affiliation[a]{Universit\'a di Trieste, Dipartimento di Fisica,\\  I-34127 Trieste, Italy}

\affiliation[b]{Istituto Nazionale di Fisica Nucleare, Sezione di Trieste,\\ I-34127 Trieste, Italy}

\affiliation[c]{Istituto Nazionale di Astrofisica - Istituto di Radioastronomia,\\ I-40129 Bologna, Italy}

\affiliation[d]{W. W. Hansen Experimental Physics Laboratory, Kavli Institute for Particle Astrophysics and Cosmology, Department of Physics and
SLAC National Accelerator Laboratory, Stanford University,\\  94305, Stanford, CA, USA}

\affiliation[e]{Istituto Nazionale di Fisica Nucleare, Sezione di Bari,\\ I-70126 Bari, Italy}

\emailAdd{giacomo.principe@ts.infn.it}

\abstract{Fast radio bursts (FRBs) are one of the most exciting new mysteries of astrophysics. Their origin is still unknown, but recent observations seems to link them to Soft Gamma Repeaters and, in particular, to magnetar giant flares (MGFs). The recent detection of a MGF at GeV energies by the \textit{Fermi} Large Area Telescope (LAT) motivated the search for GeV counterparts to the $>$100 currently known FRBs. 
Taking advantage of more than 12 years of \textit{Fermi}-LAT data, we perform a search for gamma-ray emission from all the reported repeating and non-repeating FRBs. We analyse on different-time scales the \textit{Fermi}-LAT data of each individual source separately, including a cumulative analysis on the repeating ones. In addition, we perform the first stacking analysis at GeV energies of this class of sources in order to constrain the gamma-ray properties of the FRBs that are undetected at high energies. The stacking analysis is a powerful method that allow a possible detection from below-threshold FRBs providing important information on these objects. In this talk we present the preliminary results of our study and we discuss their implications for the predictions of gamma-ray emission from this class of sources.}

\FullConference{37$^{\rm{th}}$ International Cosmic Ray Conference (ICRC 2021)\\
		July 12th -- 23rd, 2021\\
		Online -- Berlin, Germany}

\begin{document}
\maketitle

\section{Introduction}
Discovered just over a decade ago \citep{2007Sci...318..777L}, fast radio
bursts (FRBs) are one of the newest astrophysical enigmas. FRBs are bright (typical fluences of few Jy) and short-duration (few ms or less) pulses at frequencies of about 1 GHz, having usually large dispersion measures (DM; the free electron density along the line of sight at a certain distance) in excess of Galactic values \citep{2019A&ARv..27....4P}.

Since their discovery in 2007, several hundreds, as of 2021, June 15, of FRBs have been reported so far \citep[e.g.; see FRB catalogs][]{2016PASA...33...45P,2021arXiv210604352T}.
In particular, an increasing number of FRBs exhibit repeating bursts random in time, and two of them show a periodic pattern in their activity cycle, as in the case of FRB\,180916 \citep{2020Natur.582..351C} and FRB\,121102 \citep{2020MNRAS.495.3551R} for which however there is only an indication of periodicity ($\sim 3 \sigma$ significance).
No clear indications for physically different populations distinguishing repeating and non-repeating sources has been obtained so far.

Despite being a recent discovery, a great interest have been attracted by the study of FRBs. In particular, FRBs may be used as powerful probes of the intergalactic medium, and emission mechanism powering these bursts may help clarify some long-standing issues in astrophysics, including the missing baryon problem and the nature of coherent emission\citep{2014ApJ...780L..33M}.

Many attempts have recently been made for understanding their progenitors. The short duration of FRBs favours models involving compact objects such as strongly magnetized neutron stars (magnetars) and massive black holes \citep{2019ARA&A..57..417C}.
Last year, for the first time, an FRB-like event was associated with a Soft Gamma Repeater (SGR\,1935+2154) and, in particular, to a Galactic magnetar giant flare (MGF) \citep[FRB 200428,][]{2020Natur.587...59B}.

The recent detection of high-energy emission, at GeV energies, from a magnetar giant flare in the Sculptor galaxy by \citep{2021NatAs.tmp...11F} motivated the search for gamma-ray counterparts to the known FRBs.
In the last years, some high-energy counterpart searches for FRBs have been performed without any significant detection \citep[see e.g,][]{2019ApJ...879...40C,2020A&A...637A..69G,2020ApJ...893L..42T,2021arXiv210500685V}, however they were based on a few dozens of FRBs.

Thanks to over 12 years of data collected by the \textit{Fermi} Large Area Telescope (LAT), and to more than 1000 published FRBs, we aim to perform the largest and deepest systematic search for gamma-ray emission from all the reported repeating and non-repeating bursts.  
We make use of different analysis techniques to search for time coincidences between exposed FRBs and events at different timescales, ranging from few seconds to days/weeks and up to several years, with \textit{Fermi}-LAT.
In addition to the study of each individual FRB source, we perform also a stacking analysis for searching the cumulative emission from the repeating FRBs.

\section{Sample of FRBs}
For our analysis we selected published FRBs from the following resources: 
\begin{itemize}
    \item 118 events from the FRBCAT\footnote{https://frbcat.org} \citep{2016PASA...33...45P};
    \item 535 repeating and non-repeating FRBs reported in the first CHIME/FRB catalog \citep{2021arXiv210604352T};
    \item 230 bursts from the 20 repeating FRBs reported by the CHIME/FRB collaboration  (http://www.chime-frb.ca/repeaters) as of June 15, 2021, including 73 bursts from the periodic FRB\,180916;
    \item 235 bursts from FRB\,121102 collected by \citet{2020MNRAS.495.3551R}.
\end{itemize}

\begin{figure}
\centering
\includegraphics[width=14cm]{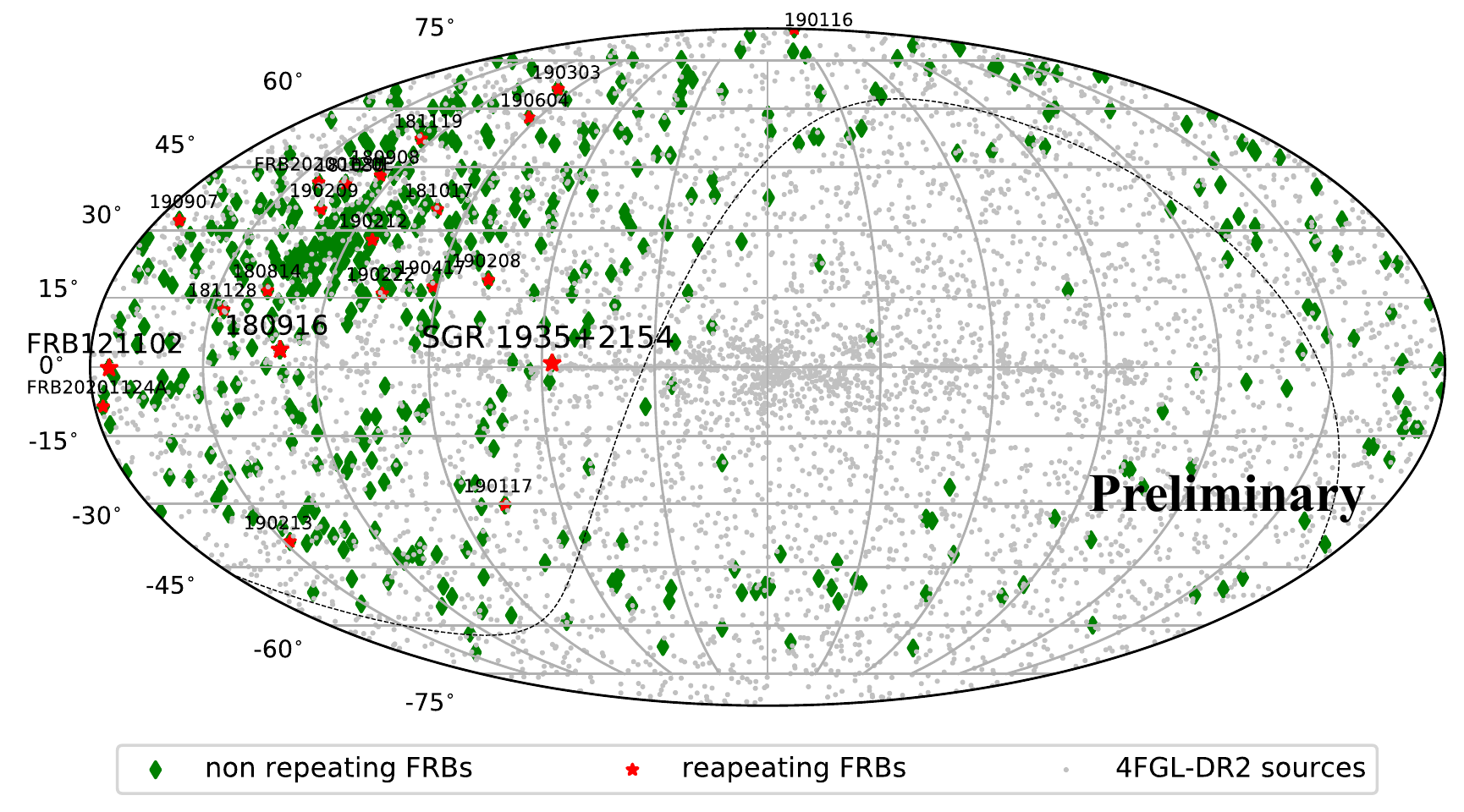}
\caption{\small \label{fig:sky_map}
Sky map, in Galactic coordinates and Mollweide projection, showing the 1025 FRBs in our sample. Repeating FRBs are labelled in the plot. All the 4FGL-DR2 sources \citep{2020ApJS..247...33A} are also plotted, with grey points, for a comparison.}
\end{figure}

Since several sources are present in more than one of the above-listed samples, we removed all the repetitions. In addition to these samples, we included the first FRB from the Galactic MGF located in SGR\,1935+2154.
Our sample consists of 1025 FRBs events, including 560 non repeating FRBs and 465 bursts from 22 repeating FRBs.
Figure \ref{fig:sky_map} shows the sky map of the selected FRBs, compared with the gamma-ray sources in the 4FGL-DR2 catalog \citep{2020ApJS..247...33A}. 
The observed asymmetric distribution between the northern and southern hemispheres, as well as the fact that all the repeating bursts are located in the northern hemisphere, is related to the large number of bursts detected by the CHIME/FRB telescope \citep{2018ApJ...863...48C}.

\begin{figure}
\centering
\includegraphics[width=0.6\columnwidth]{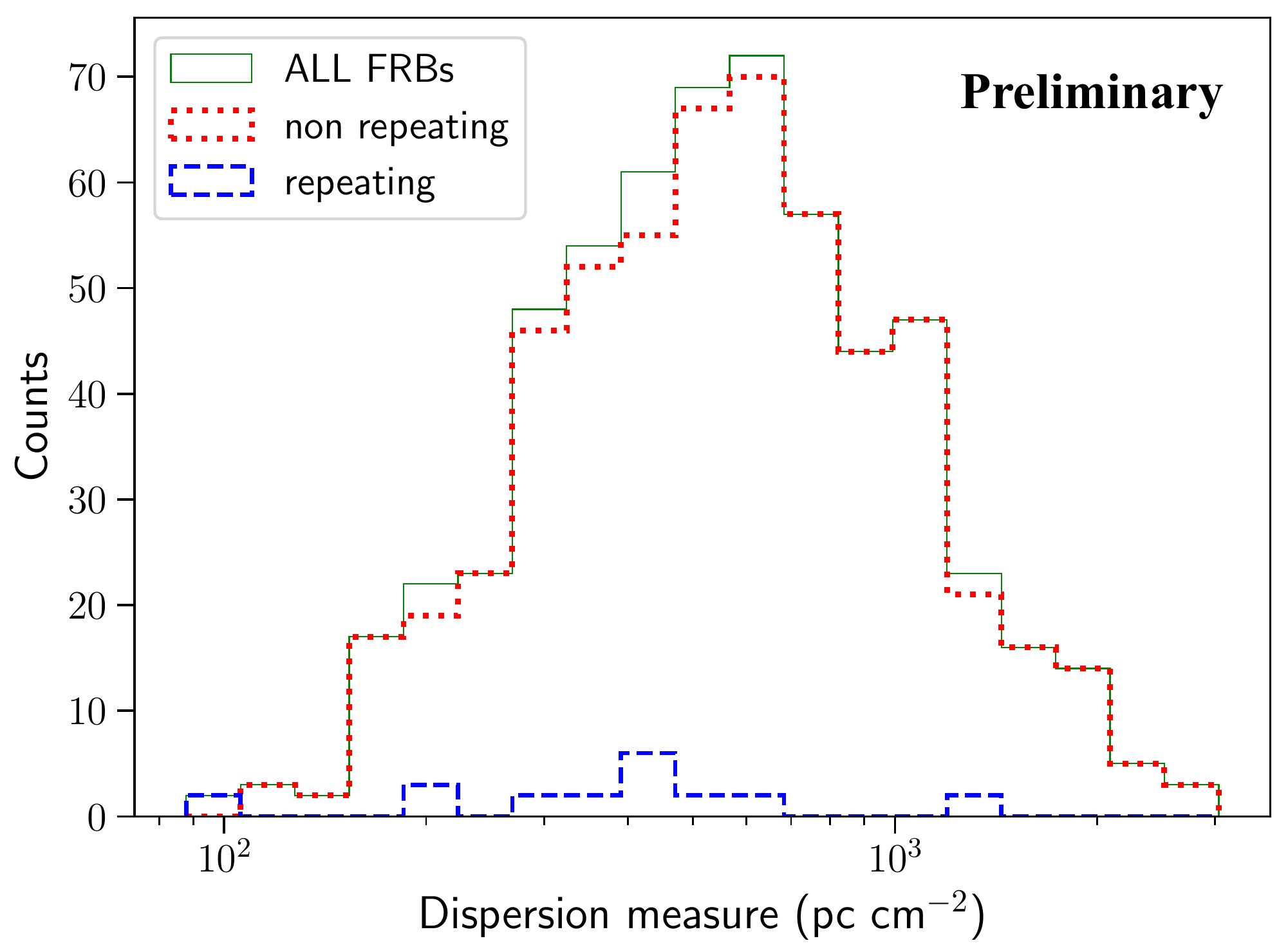}
\caption{\small \label{fig:dm_distrib}
Dispersion measure (DM) distribution of all the FRBs contained in our sample.}
\end{figure}

Figure \ref{fig:dm_distrib} shows the dispersion measure (DM) distribution of the selected FRBs. 

\noindent The large dispersion measure of the observed bursts suggests their possible extragalactic origin. To date the detection of FRBs with associated small (arcsec) error boxes has allowed the detection of fifteen putative host galaxies\footnote{http://frbhosts.org/, accessed on June 15, 2021} with a luminosity distances range from 149 Mpc to 4 Gpc, giving solid bases to their cosmological origin.
Last year, for the first time, an FRB-like event (DM=332.7206) was associated to a Galactic magnetar giant flares (MGFs), SGR 1935+2154, located at a distance of 12.5 kpc \citep[FRB 200428,][]{2020Natur.587...59B,2020Natur.587...54C}.

Among the repeating bursts, the FRB\,180916 presents a significant periodicity of 16.35-day period, with all events observed in a window of 5.4 days (active phase). While a possible (3.2$\sigma$ significance) periodic activity from FRB\,121102 have been recently reported \citep{2020MNRAS.495.3551R}.
In this proceeding we focus on the preliminary results on the search of high-energy emission from the periodic FRB\,180916, while further results on all the selected FRBs will be provided in a forthcoming publication \citep{principe_in_preparation}.

\section{\textit{Fermi}-LAT analysis}
The LAT is a gamma-ray telescope that detects photons by conversion into electron-positron pairs and has an operational energy range from 20\,MeV to $\sim$\,TeV. It comprises a high-resolution converter tracker (for direction measurement of the incident gamma rays), a CsI(Tl) crystal calorimeter (for energy measurement) and an anti-coincidence detector to identify the background of charged particles \citep{2009ApJ...697.1071A}. Thanks to its wide-field of view, it is able to cover the entire sky in $\sim 3$ hours, making it a very suitable $\gamma$-ray detector for investigating the high-energy counterpart of FRBs.


The search for gamma-ray emission from FRBs was performed using various analysis techniques on different time scales ranging from few seconds up to several years.
Subsequently we performed a stacking analysis of the events which were not significantly detected in the individual study, in order to investigate the general properties of the population of the FRB sources. A similar method was applied in \citet{2021MNRAS.507.4564P} to study another class of celestial objects.

First, we applied a photon counting analysis to study the coincidence of single or multiple events with the time and position of the FRBs in our sample.

Then, we analysed the \textit{Fermi}-LAT data of each individual source in our sample using a standard likelihood analysis in order to determine whether it is detected or not (using a Test Statistic TS\footnote{The test statistic (TS) is the logarithmic ratio of the likelihood $\mathcal{L}$ of a model with the source being at a given position in a grid to the likelihood of the model without the source, TS=$2 \log \frac{\mathcal{L}_\mathrm{src}}{\mathcal{L}_\mathrm{null}}$ \citep{1996ApJ...461..396M}.} $>25$ as a threshold).

We analysed each individual burst using time intervals ranging from few seconds to days/months centred on the arrival time of the FRB. We also analysed the 12.7 years (August 5, 2008 - April 5, 2021) of \textit{Fermi}-LAT data. This latest time-interval is motivated by the search of possible new steady sources (not present in the latest publish LAT catalog \citep{2020ApJS..247...33A}) from which the FRBs may be originated.

In addition to the study of each individual burst contained in our sample, we performed an 'ad-hoc' analysis for the repeating FRBs searching for cumulative signal from all the intervals when bursts have been reported. We used therefore a 'folding' analysis procedure, combining together 1000-seconds time intervals of all the events from a FRB source.

Furthermore, a special attention has been given to the periodic FRB\,180916, on which a folding analysis have been performed considering both the 0.6-day and 5.4-day phase windows. These time-intervals correspond to the windows on which 50\% and 100\% of the events have been detected  \citep{2020Natur.582..351C}, 
We analysed both the active phases after the FRB discovery as well as all the periods for the 12.7 years of \textit{Fermi}-LAT available. A similar approach has been applied also to FRB\,121102, for which a possible periodicity of $\sim$157 days period have been recently reported \citep{2020MNRAS.495.3551R}.

The binned likelihood analysis (which consists of model optimisation, and localisation, spectrum and variability analyses) was performed with Fermipy\footnote{http://fermipy.readthedocs.io/en/latest/} \citep{2017ICRC...35..824W}, a python package that facilitates the analysis of LAT data with the \textit{Fermi} Science Tools, of which the version 11-07-00 was used. For this analysis we selected events which have been reprocessed with the P8R3\_Source\_V2 instrument response functions (IRFs) \citep[IRFs,][]{2018arXiv181011394B}, in the energy range between 100\,MeV and 1\,TeV. The low energy threshold is motivated by the large uncertainties in the arrival directions of the photons below 100 MeV, leading to a possible confusion between point-like sources and the Galactic diffuse component. For a different analysis implementation to solve this and other issues at low energies with \textit{Fermi}-LAT see \citep{2018A&A...618A..22P}.

The analysis consists of model optimisation, source localisation and spectrum study. The counts maps were created with a pixel size of $0.1^{\circ}$. 
We selected $\gamma$-rays with zenith angle smaller than 90$^{\circ}$.
The model used to describe the ROI includes all point-like and extended LAT sources, located at a distance $<25^{\circ}$ from each target position, listed in 4FGL catalog \citep{2020ApJS..247...33A}, as well as the Galactic diffuse and isotropic emission \footnote{https://fermi.gsfc.nasa.gov/ssc/data/access/lat/BackgroundModels.html}. For the analysis we first optimised the model for the ROI, then we searched for the possible presence of new sources and finally we re-localised the source. 
During the model optimisation, we left free to vary the diffuse background and, only for the analysis on long (i.e., more than a day) time-scales, also all the spectral parameters of the sources within 5$^{\circ}$ of our targets, while only the normalisation for those at a distance between 5$^{\circ}$ and 10$^{\circ}$. 

In addition to study each individual FRB source, we perform a stacking analysis for searching the cumulative emission from the repeating FRBs, as well as on the entire population of FRB sources.

\section{Preliminary \textit{Fermi}-LAT results on FRB\,180916}
In this part we present our preliminary results on the search for high-energy emission from the periodic FRB\,180916 (z=0.0337) with \textit{Fermi}-LAT.
We analysed LAT data from 10-seconds and 1000-seconds time intervals centred on the first observed burst (MJD=58377.42972096) without finding any significant emission. 
In addition, we performed a folding analysis on the cumulative gamma-ray emission from the 73 detected bursts, using time intervals of 1000 s centred on each event, as well as on the 5.4-day active phase windows of the periodic FRB, using the 12.7 years of available LAT data.
Similarly to the analysis on short time scales, we did not find any significant gamma-ray emission also on this cumulative analysis of the periodic FRB\,180916. 

We report on Table \ref{table_ul} the 95\% upper limits on the FRB energy flux, obtained for a power-law spectral model of spectral photon index of 2.

\begin{table}
\caption{\small \label{table_ul} The 95\% upper limits on the energy flux of the periodic FRB\,180916 from 10-seconds (1) and 1000-seconds (2) time intervals centred on the first observed burst (MJD=58377.42972096), from the folding analysis of the cumulative gamma-ray emission from 1000s time intervals of the 73 detected bursts (3), as well as on the 5.4-day active phase windows (4). Upper limits were obtained for a power-law spectral model of spectral photon index of 2.}
\small
\centering
\begin{tabular}{c|c}
\hline \hline
Analysis & Upper limit on energy flux\\
 & (erg cm$^{-2}$ s$^{-1}$)\\
\hline
10 s & $7.8 \times 10^{-8}$\\
1000 s & $1.4 \times 10^{-9}$\\
folding 1000 s of all bursts & $1.7 \times 10^{-10}$\\
5.4-days active phase (12.7 years) & $2.3 \times 10^{-12}$\\
\hline
\end{tabular}
\end{table}

\begin{figure}[h]
\centering
\includegraphics[width=0.75\columnwidth]{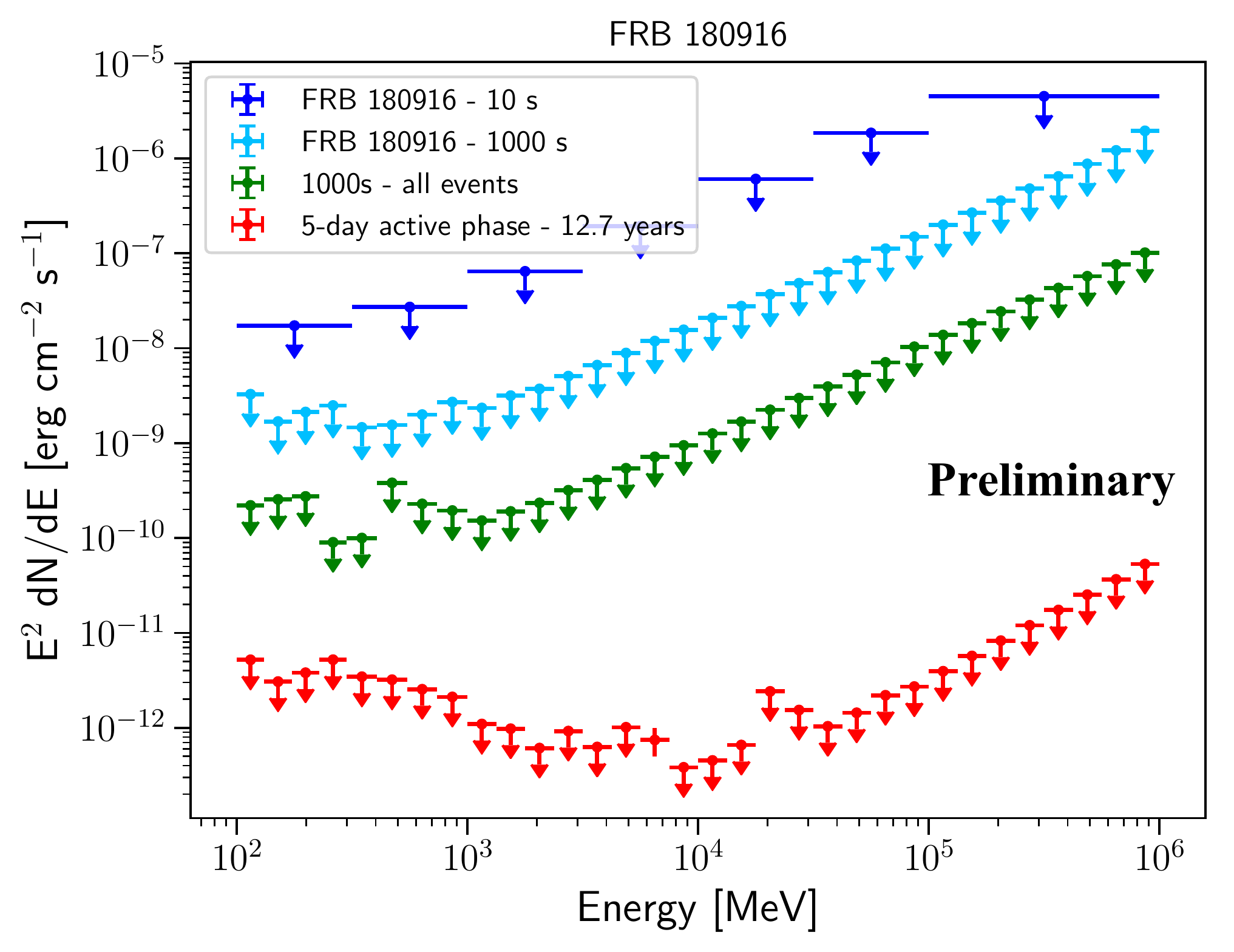}
\caption{\small \label{fig:sed_fermi} \textit{Fermi}-LAT spectrum of the periodic FRB\,180916. We plot the upper-limits for the 10 s and 1000 s time intervals centred on the first FRB event, as well as for the folding analysis on the 73 detected burst using a 1000 s time intervals, and for the 5.2-day active phase windows obtained for the 12.7 years of LAT data.
}
\end{figure}

Figure \ref{fig:sed_fermi} shows the spectral upper limits obtained on the periodic FRB\,180916 for the different time-scales used in our analysis.

\noindent Although in our analysis we did not find any detection, we provide the so-far most stringent upper limits on the gamma-ray emission from the FRB\,180916 source during its 5-day active-phase window \citep[see review][]{2021Univ....7...76N}.

\section{Conclusion}
FRBs are one of the most intriguing topics in astronomy of the last decade. We presently know of more than 1000 burst from 560 non-repeating FRBs and 22 repeating ones.
Despite their origin is still unclear, recent observations seem to associate them to Soft Gamma Repeaters and, in particular, to MGFs. Furthermore, the recent detection by the \textit{Fermi}-LAT of gamma-ray emission from a MGF located in the Sculpture galaxy (z=0.000811) \citep{2021NatAs.tmp...11F} galvanises the search of high-energy emission from FRBs. We aim to perform the largest and deepest systematic search for gamma-ray counterpart to FRBs. 

We report here the preliminary results of search for high-energy emission from the periodic FRB\,180916 (z=0.0337). Although in our analysis we did not find any detection, we provide the so-far most stringent upper limits on the gamma-ray emission from the FRB\,180916 source during its 5-day active-phase window ($L_{\gamma -ray}< 7.5 \times 10^{42}$ erg s$^{-1}$). 
Our results provide crucial information on constraining the origin of FRBs and modelling their emission mechanisms.

Further results on the search for the high-energy emission from the $\sim$1000 selected FRBs will be provided in a forthcoming paper \citep{principe_in_preparation}. 

\subsection*{ACKNOWLEDGMENTS}
We acknowledge use of the CHIME/FRB Public Database, provided at https://www.chime-frb.ca/ by the CHIME/FRB Collaboration.

The \textit{Fermi}-LAT Collaboration acknowledges support from NASA and DOE (United States), CEA/Irfu, IN2P3/CNRS, and CNES (France), ASI, INFN, and INAF (Italy), MEXT, KEK, and JAXA (Japan), and the K.A. Wallenberg Foundation, the Swedish Research Council, and the National Space Board (Sweden).

%
%
%


\begin{thebibliography}{99}
\bibitem[Lorimer et al.(2007)]{2007Sci...318..777L} Lorimer, D.~R., Bailes, M., McLaughlin, M.~A., et al.\ 2007, Science, 318, 777. 

\bibitem[Petroff et al.(2019)]{2019A&ARv..27....4P} Petroff, E., Hessels, J.~W.~T., \& Lorimer, D.~R.\ 2019, \aapr, 27, 4. 

\bibitem[Petroff et al.(2016)]{2016PASA...33...45P} Petroff, E., Barr, E.~D., Jameson, A., et al.\ 2016, \pasa, 33, e045. 

\bibitem[The CHIME/FRB Collaboration et al.(2021)]{2021arXiv210604352T} The CHIME/FRB Collaboration, :, Amiri, M., et al.\ 2021, arXiv:2106.04352

\bibitem[Chime/Frb Collaboration et al.(2020)]{2020Natur.582..351C} Chime/Frb Collaboration, Amiri, M., Andersen, B.~C., et al.\ 2020, \nat, 582, 351. 

\bibitem[Rajwade et al.(2020)]{2020MNRAS.495.3551R} Rajwade, K.~M., Mickaliger, M.~B., Stappers, B.~W., et al.\ 2020, \mnras, 495, 3551. 

\bibitem[McQuinn(2014)]{2014ApJ...780L..33M} McQuinn, M.\ 2014, \apjl, 780, L33. 

\bibitem[Cordes \& Chatterjee(2019)]{2019ARA&A..57..417C} Cordes, J.~M. \& Chatterjee, S.\ 2019, \araa, 57, 417. 

\bibitem[Bochenek et al.(2020)]{2020Natur.587...59B} Bochenek, C.~D., Ravi, V., Belov, K.~V., et al.\ 2020, \nat, 587, 59. 

\bibitem[Fermi-LAT Collaboration et al.(2021)]{2021NatAs.tmp...11F} Fermi-LAT Collaboration, Ajello, M., Atwood, W.~B., et al.\ 2021, Nature Astronomy. 

\bibitem[Cunningham et al.(2019)]{2019ApJ...879...40C} Cunningham, V., Cenko, S.~B., Burns, E., et al.\ 2019, \apj, 879, 40. 

\bibitem[Guidorzi et al.(2020)]{2020A&A...637A..69G} Guidorzi, C., Marongiu, M., Martone, R., et al.\ 2020, \aap, 637, A69. 

\bibitem[Verrecchia et al.(2021)]{2021arXiv210500685V} Verrecchia, F., Casentini, C., Tavani, M., et al.\ 2021, arXiv:2105.00685

\bibitem[Tavani et al.(2020)]{2020ApJ...893L..42T} Tavani, M., Verrecchia, F., Casentini, C., et al.\ 2020, \apjl, 893, L42. 


\bibitem[Abdollahi et al.(2020)]{2020ApJS..247...33A} Abdollahi, S., Acero, F., Ackermann, M., et al.\ 2020, \apjs, 247, 33. 

\bibitem[CHIME/FRB Collaboration et al.(2018)]{2018ApJ...863...48C} CHIME/FRB Collaboration, Amiri, M., Bandura, K., et al.\ 2018, \apj, 863, 48.


\bibitem[CHIME/FRB Collaboration et al.(2020)]{2020Natur.587...54C} CHIME/FRB Collaboration, Andersen, B.~C., Bandura, K.~M., et al.\ 2020, \nat, 587, 54. 

\bibitem[Principe et al.(submitted)]{principe_in_preparation} Principe, G., Omodei N., Di Venere L., et al.\ in preparation.

\bibitem[Atwood et al.(2009)]{2009ApJ...697.1071A} Atwood, W.~B., Abdo, A.~A., Ackermann, M., et al.\ 2009, \apj, 697, 1071.

\bibitem[Principe et al.(2021)]{2021MNRAS.507.4564P} Principe, G., Di Venere, L., Orienti, M., et al.\ 2021, \mnras, 507, 4564. doi:10.1093/mnras/stab2357


\bibitem[Mattox et al.(1996)]{1996ApJ...461..396M} Mattox, J.~R., Bertsch, D.~L., Chiang, J., et al.\ 1996, \apj, 461, 396.

\bibitem[Wood et al.(2017)]{2017ICRC...35..824W} Wood, M., Caputo, R., Charles, E., et al.\ 2017, 35th International Cosmic Ray Conference (ICRC2017), 301, 824

\bibitem[Bruel et al.(2018)]{2018arXiv181011394B} Bruel, P., Burnett, T.~H., Digel, S.~W., et al.\ 2018, arXiv:1810.11394

\bibitem[Principe et al.(2018)]{2018A&A...618A..22P} Principe, G., Malyshev, D., Ballet, J., et al.\ 2018, \aap, 618, A22. 

\bibitem[Nicastro et al.(2021)]{2021Univ....7...76N} Nicastro, L., Guidorzi, C., Palazzi, E., et al.\ 2021, Universe, 7, 76. 




\end{thebibliography}
\end{document}